\RequirePackage{ifpdf}
\documentclass[12pt]{JHEP3}
\pdfoutput = 1

\usepackage{color}
\usepackage{mathtools}
\usepackage{multicol}
\usepackage{braket} 
\usepackage{epsfig,bm}
\epsfclipon
\usepackage{amsmath,amssymb,amsbsy,amstext,amsthm}

\newcommand{\roughly}[1]{\mathrel{\raise.3ex\hbox{$#1$\kern-0.85em
\lower1ex\hbox{$\sim$}}}}

\newcommand{\lsim}{\roughly<}
\newcommand{\gsim}{\roughly>}

\usepackage{ulem}

\theoremstyle{definition}

\title{Theoretical implications of detecting gravitational waves}

\author{Ghazal Geshnizjani${}^{1,2}$, William H. Kinney${}^{3}$\\
$^1$Department of Applied Mathematics, University of Waterloo, Waterloo, Ontario, N2L 3G1, Canada\\ $^2$Perimeter Institute for Theoretical Physics, 31 Caroline St. N., Waterloo, ON, N2L 2Y5, Canada\\$^3$Department of Physics, SUNY at Buffalo, Buffalo, NY 14260, United States of America;\\
 }
\date{}

\abstract{This paper is the third in a series of theorems which state how cosmological observations can provide evidence for an early phase of acceleration in the universe. It was demonstrated in \cite{Geshnizjani:2011dk,Geshnizjani:2013lza}, that the observed power spectrum for scalar perturbations forces all possible alternative theories of inflation to theories other than General Relativity. It was shown that generically, without a phase of accelerated expansion, these alternatives have to break at least one of the following tenets of classical general relativity: the Null Energy Condition (NEC), subluminal signal propagation, or sub-Planckian energy densities. In this paper we prove how detection of primordial gravitational waves at large scales can provide independent evidence to support a phase of accelerated expansion. This proof does not rely on the spectral index for tensor modes but relies on validity of quantum field theory in curved space time and tensor modes being sourced from adiabatic vacuum fluctuations.  Our approach, like in the case of scalars, is proof by contradiction: we investigate the possibility of a detectable tensor signal sourced by vacuum fluctuations in a non-accelerating, sub-Planckian universe using cosmological perturbation theory and derive contradictory limits on cosmological dynamics. The contradiction implies that one or more of our axioms for early universe must have been broken. The bound from tensor perturbations is not only independent of, but also stronger than the one obtained from scalar power spectrum.}

\preprint{}

\begin{document}
\section{Introduction}
Observation of primordial gravitational waves is one of the cosmological  enterprises that, if successful, will open a new window to beginning of our universe and will also  provide evidence for quantized gravitational effects, {\it i.e}. gravitons.
The recent detection of odd parity, or ``B-mode'' polarization by the BICEP2 telescope \cite{Ade:2014xna} suggests the possibility that there may be a detectable signal from primordial gravitational waves produced in the very early universe. Subsequent analysis of multi-frequency data from Planck indicates that the B-mode signal in the BICEP2 field is at least partly due to galactic dust \cite{Adam:2014oea}, but the potential remains for near-future Cosmic Microwave Background (CMB) measurements to detect B-mode polarizaton from primordial gravitational waves, also referred to as tensor modes. 

In this paper, we investigate the consequence for early-universe physics of such a detection. Primordial tensor modes are a well-known prediction of many cosmological inflationary models \cite{Guth:1980zm,Linde:1981mu,Albrecht:1982wi,Kazanas:1980tx,Starobinsky:1980te,Sato:1981ds,Sato:1980yn,Mukhanov:1981xt,Guth:1982ec,Hawking:1982cz,Bardeen:1983qw,Linde:1983gd}, and their detection is often asserted to be a ``smoking gun'' for inflation in the early universe. In this paper, we evaluate the generality of this claim in a systematic way by examining  the generation of tensor perturbations from vacuum fluctuations in general Friedmann-Robertson-Walker (FRW) backgounds. We derive general conditions for the production of tensor modes at large cosmological scales with amplitudes detectable by current or future measurements. 

In previous work, we have examined the consequences of the observed spectrum of {\it scalar} modes, or density perturbations for early universe cosmology. We showed that, assuming standard General Relativity, to generate scalar perturbations consistent with observation,  at least one of the following four conditions is necessary (although not necessarily sufficient) \cite{Geshnizjani:2011dk,Geshnizjani:2013lza}:
\begin{enumerate}
\item {Accelerated expansion.} \footnote{Note this is a weaker condition than requiring a phase of slow-roll inflation.}
\item {Superluminal signal propagation, for example a speed of sound $c_S > c$.}
\item {Violation of the Null Energy Condition (NEC), for example in a regular bouncing cosmology.}
\item {Energy density higher than the Planck density, indicating a breakdown of standard physics and the need for a quantum gravitational model. A singular bouncing cosmology is an example of that.}
\end{enumerate}
Note that only the first condition is consistent within general relativity and the remaining three point out to the break down of semi classical gravity. 
Scalar perturbations in cosmology are complicated by the fact that the effective mass of the {\it Mukhanov-Sasaki} variable $v$ (related to the gauge-invariant scalar $\zeta$), depends on the product of the time evolution of the cosmological background and its derivatives in an involved way and they also depend on speed of sound of the cosmological fluid. Thus, the presence of superhorizon scalar perturbations in the CMB does not {\it per se} require early-universe inflation. However, primordial gravitational waves are a simpler and cleaner probe of early-universe expansion, since they do not couple directly to the cosmological fluid: the quantum generation of tensor modes is only sensitive to the overall dynamics of the background spacetime.  Thus, it is reasonable to expect that an observable spectrum of superhorizon tensor modes will provide a much stronger constraint on the physics of the early universe. 

In this paper we examine the general consequences of a detectably large spectrum of superhorizon primordial tensor modes in the CMB. We  show that inflation is not, in general, required to enhance vacuum tensor fluctuations on superhorizon scales. For example, a spectrum of tensor modes on scales larger than the Hubble length is generically created in a phase transition between radiation domination and matter domination in the early universe. Such a spectrum, however, is strongly scale-dependent, and is  too small to result in an observable signal.  We find that,  a tensor signal with amplitude $P_H \sim 10^{-10}$, as suggested by the BICEP2 measurement, in non-accelerating expansion histories is not consistent within the framework of standard General Relativity. Given that scenarios that have non-expanding expansion in the past have to violate general relativity, we first prove a similar theorem derived for scalars power-spectrum for tensor modes considering only an {\it expanding} universe (consistent with General Relativity). We demonstrate  that an observable tensor signal can only be produced either by inflation, or by intrinsically quantum-gravitational effects. We also observe that our argument can be applied to large class of other scenarios which do not fall into the standard General Relativity paradigm, such as a {\it contracting} universe, {\it Horndeski} theories and {Horava-Lifshitz} gravity.

The paper is organized as follows: In Section \ref{sec:WKB} we discuss the quantum generation of tensor perturbations in a FRW background. In Section \ref{sec:Bounds} we discuss general bounds on cosmological evolution consistent with an observably large tensor spectrum. In Section \ref{sec:Alternatives} we discuss bouncing cosmology and the possibility of modifications to General Relativity. In Section \ref{sec:Conclusions} we present a summary and conclusions. 


\section{Gravitational waves in an expanding background}
\label{sec:WKB}

In this section we derive bounds on the production of gravitational waves in an expanding cosmological background in the absence of inflation. The second-order action for tensor metric perturbations (assuming standard General Relativity) is\footnote{The prime symbol denotes derivative with respect to conformal time $\tau$, while the dot symbol refers to derivative with respect to cosmological time $t$.}:
\begin{equation}\label{action}
S_2={1\over 2}\int d^3x~ d\tau \left [ v^{\prime2}- (\nabla v)^2 +{a''\over a}v^2\right ], 
\end{equation}
where, 
\begin{equation}
v\equiv M_{pl} ~a~h_{+/\times}.
\end{equation}
The equation of motion for $v$ in Fourier space is then  given by 
\begin{equation}\label{eqnforvk}
v_k''+\left(k^2-{a''\over a}\right)v_k=0. 
\end{equation}
Note that 
\begin{eqnarray}\label{mass}
{a''\over a}=a^2H^2(2-\epsilon), 
\end{eqnarray}
where $\epsilon\equiv -\dot{H}/H^2$ determines the equation of state
\begin{equation}
\epsilon = \frac{3}{2} \left(1 + \frac{p}{\rho}\right). 
\end{equation}
We therefore have accelerating expansion $\ddot a > 0$ ({\it i.e.} inflation) for $\epsilon < 1$, and decelerating expansion for $\epsilon > 1$. 
For $\epsilon \neq 2$, in the limit where $k^2\ll {a''/ a}$, the above equation admits the following general solutions:
\begin{equation}\label{freezedsoliution}
v_k=C_1 a(\tau)+C_2 a(\tau) \int {d\tau\over a^2(\tau)},
\end{equation}
which when divided by $a(\tau)$ leads to a constant mode and a time-dependent mode for $h$. Based on Eq. \ref{mass}, we define a transition scale, the {\it freezing} radius such that $a''/a=1/R_{fr}^2$. This scale identifies the scale above which the mode equation is no longer a system of simple harmonic oscillator
\begin{equation}\label{RfRH}
R_{fr}= {R_H\over \sqrt{|2-\epsilon|}},
\end{equation}
where $R_H\equiv 1/|aH|$ refers to comoving Hubble radius. As we will show in few paragraphs this is the scale where WKB approximation breaks down. For modes with wavelength larger than the freezing radius, or equivalently wavenumber $k \ll 1/R_{fr}$, the $v_k \propto a$ mode is ``frozen'', with
\begin{equation}
\frac{v_k}{a} \rightarrow {\rm const.}
\end{equation}
The time-dependent mode 
\begin{equation}
\frac{v_k}{a} \propto \int{\frac{d \tau}{a^2(\tau)}}, 
\end{equation}
can be either decaying or growing, depending on equation of state. 
For slowly varying equation of state, $\epsilon \sim {\rm const.}$, 
\begin{equation}
a \propto t^{1/\epsilon} \propto \tau^{1 / (\epsilon - 1)},
\end{equation}
so that 
\begin{equation}
\frac{v_k}{a} \propto \int{\frac{d \tau}{a^2(\tau)}} = \int{\tau^{2 / (1 - \epsilon)} d\tau} \propto \tau^{(3 - \epsilon) / (1 - \epsilon)},
\end{equation}
which is decaying for $\epsilon < 3$ and growing for $\epsilon > 3$\footnote{This mode remains decaying for $\epsilon<1$, since in that case $|\tau|$ is decreasing in time.}. We will neglect the time-dependent mode in the discussion laid out in the main body of the paper and refer readers to Appendix \ref{growing} to see how our argument is applicable to scenarios with a growing mode such as those with stiff equation of state, $\epsilon \gg 3$, and why that does not change our main conclusion.

Mode freezing will take place as long as the freezing horizon {\it shrinks} in comoving units. Defining the number of e-folds\footnote{We are setting the value of present day scale factor to one.} $N\equiv \ln{a}$, in an expanding background $d N > 0$ and the change in $R_{fr}$ is 
\begin{equation}\label{eqnfreezeout}
\frac{d \ln{R_{fr}}}{dN} = -1+\epsilon + \frac{1}{2( 2-\epsilon)} \frac{d\epsilon}{dN}. 
\end{equation}
Mode freezing occurs for $d \ln{R_{fr}} < 0$. For constant equation of state, this can only happen via inflationary expansion,
\begin{equation} \label{shrinkR}
\frac{d \ln{R_{fr}}}{dN} = -1+\epsilon < 0\ \Rightarrow \epsilon < 1.
\end{equation}
However, it is clear from (\ref{eqnfreezeout}) that mode freezing can also occur for rapidly varying equation of state\footnote{Note that for both of these inequalities to hold we assumed $H>0$ and universe has only been in an expanding phase, otherwise these inequalities are reversed.},\
\begin{eqnarray} \label{changeinepsilon}
\frac{d \epsilon}{d N} &<& 2 \left(1 - \epsilon\right) \left(2 - \epsilon\right)<0\quad \ (\epsilon < 2)\cr
\frac{d \epsilon}{d N} &>& 2 \left(1 - \epsilon\right) \left(2 - \epsilon\right)>0\quad \ (\epsilon > 2).
\end{eqnarray}
A simple example is a phase transition between radiation domination ($\epsilon \simeq 2$) and matter domination, ($\epsilon \simeq 3/2$). For $\epsilon \rightarrow 2$, the freezing horizon diverges\cite{Powell:2006yg},
\begin{equation}
R_{fr}^{-2} = \frac{a''}{a} \simeq 0.
\end{equation}
In this limit, the mode equation (\ref{eqnforvk}) reduces to a pure Klein-Gordon form, which can admit a normalized adiabatic vacuum\footnote{Adiabatic vacuum is analogous to Bunch-Davis initial condition for none  de Sitter spacetimes} solution at all scales
\begin{equation}
v_k = \frac{1}{\sqrt{2 k}} e^{-i k \tau}.
\end{equation}
The tensor power in the vacuum state is then
\begin{eqnarray}
{\cal P}^h_k&\equiv& k^3{ |v_k^2|\over M_{Pl}^2 a^2} \\&\approx& \frac{1}{2}\left(\frac{k}{M_p a}\right)^2.
\end{eqnarray}
Consider a rapid phase transition from the radiation-dominated state to a matter-dominated state at scale factor $a = a_{fr}$. Then the freezing horizon shrinks from infinite size to of order the Hubble length,
\begin{equation}
R_{fr} = \sqrt{2} R_h.
\end{equation}
and the tensor power spectrum will freeze out with amplitude
\begin{equation}
{\cal P}^h_k \approx  \frac{1}{2}\left(\frac{k}{M_p a_{fr}}\right)^2,
\end{equation}
producing a strongly blue spectrum of tensor modes $P\left(k\right) \propto k^2$, with amplitude suppressed by factors of the Planck mass. This mechanism provides a way to produce a blue tensor spectrum in an expanding universe. 

It is straightforward to generalize the simple analysis above to a more general WKB solution as follows: In the case of an adiabatic vacuum, the leading WKB condition\footnote{The existence of a WKB regime and solution admitting adiabatic initial conditions, prohibits the function $R_{fr}(\tau)$ from having sharp spikes} reduces to 
\begin{equation}
\omega_k^2 \gg \left|{\omega_k'' \over 2\omega_k}-{3\omega_k'^2\over 4\omega_k^2}\right|,
\end{equation}
where $\omega_k^2 \equiv k^2-a''/a~$. In the regime that the WKB approximation holds, the WKB solution is
\begin{equation}\label{vacuum}
v_k={1\over \sqrt{ 2 \omega_k}}\exp\left[-i \int^\tau \omega_k(\tau)d\tau\right].
\end{equation}
This regime for smooth functions of $R_{fr}(\tau)$ coincides with $k >> 1/R_{fr}$. In this limit the power spectrum takes the same form as for radiation domination,
\begin{eqnarray}
{\cal P}^h_k \equiv k^3{ |v_k^2|\over M_{Pl}^2 a^2} &\approx& \frac{1}{2} \left(\frac{k}{M_p a}\right)^2\label{powerk}\\
&\approx& 2 \pi^2\left[\frac{l_{pl}}{\lambda}\right]^2(1+z)^2\label{amplk}
\end{eqnarray}
 Note that  $k$ and $\lambda$ are the comoving wave number and wavelength. This relation demonstrates the importance of the freezing mechanism at early times in order to produce a detectable primordial signal (the sensitivities of the current experiments are $10^{-10}$). If the cosmology is such that modes do not exit the adiabatic regime before the {\it Last Scattering} of CMB photons ($z_{LS}$) then the tensor power spectrum for any cosmological wavelengths ($\lambda\gg 10^9 l_{pl}$) would remain unobservable. For example, if we substitute the wavelength, $\lambda\sim 580 Mpc$ (corresponding to $l\sim 150$) in above relation we obtain 
\begin{equation}\label{Tvacfluc}
{\cal P}^h_k\sim 2.5 \times 10^{-119}(1+z_{LS})^2
\end{equation}
On the other hand, the freezing mechanism at some redshift $z_{fr}>z_{LS}$ can preserve the power for these modes at larger amplitudes
 \begin{equation}\label{Tvacfluc2}
 {\cal P}^h_k\sim 2.5\times 10^{-119}(1+z_{fr})^2.
 \end{equation}
As is evident from above equations, the freezing transition must precede even the Big Bang Nucleosynthesis (BBN) ($z\sim 10^9$) era by many orders of magnitude in redshift to lead to any observable tensor spectrum. We come back to this point in next section.
 
In the next section, we derive bounds on the amplitude of the tensor spectrum, and show that an observably large amplitude is inconsistent with cosmological evolution with energy density below the Planck density. 


\section{Bounds on the tensor amplitude in a non-inflationary universe}
\label{sec:Bounds}

In this section, we derive two independent bounds on the tensor amplitude in a non-inflationary background which show that an observable tensor component is inconsistent with sub-Planckian cosmological energy density during mode freezing. We first assume that inflation did not occur at any time in the early universe, so that $\epsilon > 1$ and the cosmological expansion was decelerating at all times. This corresponds to a monotonically {\it growing} comoving Hubble length,
\begin{equation}\label{eqnComovingHubble}
\frac{d R_H}{d N} = - \frac{1}{a H} \frac{d (a H)}{d N} = \epsilon - 1 > 0. 
\end{equation}
It is also easy to trace back the size of the wavelengths corresponding to present large-scale observations and see that they were far outside of the  Hubble Radius at BBN. Therefore, to create tensor perturbations on these scales, we must have a freezing length much greater than the Hubble length at the freezing transition,
\begin{equation}
R_{fr} = \frac{R_H}{\sqrt{2 - \epsilon}} \gg R_H.
\end{equation}
Satisfying this inequality requires an equation of state extremely close to radiation domination, $\epsilon \simeq 2$. Tensor modes will be produced via mode freezing on super-Hubble scales provided that variations in $\epsilon$ satisfy (\ref{changeinepsilon}). Mode freezing occurs when the comoving wavelength is approximately the same size as the freezing horizon, or
\begin{equation} \label{kfrRfr}
\frac{k}{a_{fr}(k)} \simeq \frac{1}{a_{fr}R_{fr}} \simeq H \sqrt{2 - \epsilon_{fr}}.
\end{equation}
 will have amplitude
\begin{equation}\label{eqntensamp}
 {\cal P}^h_k \approx  \frac{1}{2}\left(\frac{k}{M_P a_{fr}(k)}\right)^2 \simeq  \frac{1}{2} \frac{H_{fr}^2}{M_P^2} \left\vert 2 - \epsilon_{fr}\right\vert. 
\end{equation}
For the tensors produced to be observable by current experiments, we must have 
\begin{equation}
 {\cal P}^h_k \gsim 10^{-10}.
\end{equation}
In inflation, this is easily accomplished, since for $\epsilon \ll 1$,
\begin{equation}
 {\cal P}^h_k \simeq \frac{H^2}{M_P^2} \simeq 10^{-10}\ \Rightarrow H \sim 10^{-5} M_P \sim 10^{14}\ {\rm GeV}.
\end{equation}

However, close to radiation-dominated expansion, $\epsilon \sim 2$, would require a much higher Hubble scale. Consider a wavenumber $k$ within the range consistent with the BICEP2 detection of B-mode CMB polarization \cite{Ade:2014xna},
\begin{equation}
30 < \ell < 150.
\end{equation}
Using\footnote{We are setting the comoving distance to LSS, $D_A$, to $14 Gpc$\cite{Ade:2013zuv} and assuming $\ell\sim D_A k$.} $\ell \simeq 10000 h^{-1} k$,
\begin{equation}
0.003\ {\rm h Mpc}^{-1} < k < 0.015\ {\rm h Mpc}^{-1},
\end{equation}
or, in Planck units,
\begin{equation}\label{eqnkInPlanck}
1.2 \times 10^{-60} < \frac{k}{M_P} < 6.1 \times 10^{-60}.
\end{equation}
The expansion history of the universe is strongly constrained for redshifts up to the epoch of Big-Bang Nucleosynthesis (BBN), and largely unconstrained for higher redshifts. However, if we assume non-inflationary expansion, we must have that the comoving Hubble length $R_H = (a H)^{-1}$ was growing with expansion at all times, so that the Hubble length was smaller than its value at BBN at all times before BBN,
\begin{equation}
R_H(z > z_{\rm BBN}) < R_H(z_{\rm BBN}).
\end{equation}
Combining this inequality with Eq. (\ref{kfrRfr}), for a given wavenumber $k$, we have an inequality
\begin{equation} \label{kahfr}
\left(\frac{k}{(a H)_{fr}}\right) = \sqrt{2 - \epsilon_{fr}} < \left(\frac{k}{(a H)_{\rm BBN}}\right).
\end{equation}
We can therefore use the ratio of the wavenumber to the Hubble constant at the epoch of BBN to constrain the equation of state $\epsilon$ at freezeout. From (\ref{eqnComovingHubble}), we have for $\epsilon = {\rm const.}$
\begin{equation}
a H \propto (1 + z)^{\epsilon - 1},
\end{equation}
so that for radiation domination, $a H \propto (1+z)$, and for matter domination $a H \propto \sqrt{1 + z}$. Then
\begin{equation}
(a H)_{\rm BBN} =  (a H)_0 \sqrt{1 + z_{\rm eq}} \left(\frac{1 + z_{\rm BBN}}{1 + z_{\rm eq}}\right) \sim 6000\ {\rm h Mpc^{-1}},
\end{equation}
where $z_{\rm eq} \simeq 3 \times 10^{3}$ and $z_{\rm BBN} \sim 10^{9}$ are the redshifts of matter/radiation equality and BBN, respectively, and we are ignoring Dark Energy. Then for $k \sim 0.015\ {\rm h Mpc}^{-1}$, we have a bound on $\epsilon_{fr}$,
\begin{equation}  \label{epsilon150}
\sqrt{2 - \epsilon_{fr}} < 2.5 \times 10^{-6}, 
\end{equation}
while for $k \sim 0.003\ {\rm h Mpc}^{-1}$,
\begin{equation} \label{epsilon30}
\sqrt{2 - \epsilon_{fr}} < 5 \times 10^{-7}, 
\end{equation}
From Eq. (\ref{eqntensamp}), we then have a {\it lower bound} on $H_{fr}$ in Planck units
\begin{eqnarray}\label{eqnBoundOnH}
\frac{H_{fr}}{M_P} &>& \frac{10^{-5}}{\sqrt{2 - \epsilon_{fr}}} \gsim 4 \quad\quad for \quad  \ell =150 \nonumber\\
\frac{H_{fr}}{M_P} &>& \frac{10^{-5}}{\sqrt{2 - \epsilon_{fr}}} \gsim 20 \quad\quad for \quad \ell =30  
\end{eqnarray}
Therefore, we must have energy density at freezeout
\begin{equation}
\frac{\rho_{fr}}{M_P^4} \sim \left(\frac{H_{fr}}{M_P}\right)^2 > 16-400 \quad\quad for \quad \ell \in  (30,150) . 
\end{equation}
Note that we have assumed that observable tensor modes require ${\cal P}^h_k \geq 10^{-10}$. From Eq. (\ref{eqnBoundOnH}), we see that we could relax this by two orders of magnitude and still require a Hubble parameter close to the Planck scale, ${\cal P}^h_k \simeq 10^{-12}\ \Rightarrow H \sim M_P$. We therefore conclude that in {\it an expanding universe}, observable tensor fluctuations require one of two conditions:
\begin{itemize}
\item Inflation: $\epsilon < 1$
\item{Super-Planckian energy density, $\rho \gsim M_P^4$.}
\end{itemize}
This is the main result of this paper. 

A complementary bound can be obtained directly from the wavelength of the tensor modes today and the amount of expansion since BBN, as follows: From Eqs. (\ref{amplk}) we know once the freezing transition occurs
\begin{eqnarray}\label{ampfr}
{\cal P}^h_k  &\approx&  \frac{1}{2} \left(\frac{k}{M_p a_{fr}}\right)^2.
\end{eqnarray}
Substituting $k$ from (\ref{eqnkInPlanck}) and requiring ${\cal P}^h_k\gsim 10^{-10}$, we can derive an upper bound on the the ratio of scale factor at freezeout $a_{fr}$ to its present value (or a lower bound on the redshift at freezing transition),
\begin{equation}
\frac{1}{a_{fr}} \gsim 10^{-5} \left(\frac{M_P}{k}\right) \sim 1.6 \times 10^{54},
\end{equation}
which translates into a lower bound on the number of e-folds of expansion between mode freezing and today,
\begin{equation}\label{eqNLB}
\Delta N_{\rm tot} = \ln{(1 / a_{fr})} \gsim 124.8.
\end{equation}
However, we also have continuity equation,
\begin{equation}
\frac{\dot\rho}{\rho} = - 2 \epsilon H.
\end{equation}
We can therefore write 
\begin{equation}
\ln\left(\frac{\rho}{\rho_{BBN}}\right) = 2 \int{\epsilon H dt} = 2 \int{\epsilon dN} \geq 2 \epsilon_{min} \Delta N_{fr},
\end{equation}
where $\epsilon_{min}$ is the minimum value of $\epsilon$ during the evolution of the universe between freezeout and BBN. Taking non-inflationary expansion, $\epsilon_{min} > 1$, we have an upper bound on the number of e-folds between freezeout and BBN consistent with sub-Planckian energy density,
\begin{equation}
\Delta N_{fr} < \frac{1}{2} \ln\left(\frac{M_P^4}{\rho_{BBN}}\right) \sim 101. 
\end{equation}
Since we have $\Delta N_{\rm BBN} \sim \ln{10^9} \sim 21$ e-folds of expansion since BBN, we can write an upper bound on the number of e-folds since freezeout
\begin{equation}\label{eqNUB}
\Delta N_{\rm tot} < \Delta N_{fr} + \Delta N_{\rm BBN} \simeq 122.
\end{equation}
Comparing this with our lower bound (\ref{eqNLB}) from the perturbation amplitude, we see again that sufficiently large perturbations require energy densities at freezeout of order the Planck density or larger. 

To close this section, let us remind readers that in this section we assumed the time-dependent mode of the tensor perturbations remains sub-dominant after the freezing transition. We refer interested readers to Appendix \ref{growing} to show why this is a reasonable assumption. This concludes the proof of the theorem.

\section{Provisions for modified gravity, quantum gravity inspired scenarios or other sources for gravitational waves}
\label{sec:Alternatives}

As we noted above, our proof shows that an observable tensor spectrum requires that one or more of the theoretical axioms we have made must be broken. A simple and convenient conclusion would be universe having gone through an accelerated phase without breaking general relativity. That would allow for both accessing a large number of e-foldings as well as the freezing transition. However, it could just as well be that local or non-local quantum effects beyond standard effective field theory such as string theory or loop quantum gravity led to violation of NEC, imposed a different initial condition, or modified the gravitational action and continuity equation significantly. Another less exotic possibility is that topological defects such as cosmic strings gave rise to production of tensor modes. The latter possibility has already been investigated in Refs. \cite{Lizarraga:2014eaa,Moss:2014cra}. A generic feature of the B-mode polarization power spectrum for defects seems to be that it grows towards smaller scales. Therefore, a detection at large scales implies a substantial contribution at higher multipoles which are already constrained by CMB observations. 

As for quantum gravitational effects, as long as the theory of quantum gravity itself is not established, anything we say is at best speculative. Here we just comment about some possible manifestations of quantum effects.


\subsection{Contracting scenarios}

Many of alternatives of Inflation assume that the perturbations were generated in a contracting phase. This implies universe has undergone a bounce prior to the expanding phase (\cite{Battefeld:2014uga} provides a comprehensive reviews of these models). Some scenarios consider the bounce to be ``singular'' such that energy density diverges towards the bounce. 
These models assume semi-classical gravity is a valid description to generate perturbations in some pre-bounce phase where energy densities are sub-Planckian. As the energy crosses the Planckian bound, General Relativity can no longer be applied, but the quantum effects are such that the frozen large scale modes remain unaffected (see \cite{Lehners:2010fy} and references in there). Therefore, the amplitude at some pre-bounce time, $t_{-}^*$ in contracting regime is matched to the same value at some after-bounce time, $t_{+}^{*}$, in expanding regime. Our argument from section \ref{sec:Bounds} can be extended to these scenarios to show why they can not produce detectable signal. Note that the mode equation (Eq. \ref{eqnfreezeout}) is still applicable to a sup-Planckian pre-bounce phase. Therefore, the amplitude of vacuum fluctuation is calculated as before (Eq. \ref{powerk}) to be
\begin{eqnarray}
{\cal P}^h_k \equiv k^3{ |v_k^2|\over M_{Pl}^2 a^2} &\approx&  \frac{1}{2}\left(\frac{k}{M_p a}\right)^2.
\end{eqnarray}
The difference is that one has to substitute pre-bounce values of the scale factor in this equation. Furthermore, once again to preserve this amplitude, the amplitude must be frozen or as we argued before it will be negligible in present time.
The problem is that, again the total number of e-foldings from freezing to BBN has to be high in order for the amplitude of vacuum fluctuations be large at the freezing time (Eq.(\ref{ampfr}) and Eq. (\ref{eqNLB})). However, in the contracting phase as we go further back in time, scale factor increases and universe loses e-folds. Note that we need to satisfy 
\begin{eqnarray}
\Delta N_{tot}&= &\ln {a_{fr}\over a_{0}} \nonumber\\
&=& \ln {a_{fr}\over a_{-*}} + \ln {a_{+*}\over a_{-*}}+ \ln {a_{0}\over a_{+*}}\nonumber\\
&=& \Delta N_{ctr}+\Delta N_{QG}+\Delta N_{exp}\gsim 124.8,
\end{eqnarray}
where $ctr$, $QG$ and $exp$ refer to contracting, quantum gravity and expanding phases. Without acceleration, $\Delta N_{exp}\lsim 122$ (Eq. \ref{eqNUB}) and $\Delta N_{ctr}<0$, so the above condition requires universe to inflate even more during the quantum gravity phase than is necessary in an inflationary scenario with only an expanding history. 

Another class of models are regular bounce models \cite{Brandenberger:2012zb} where the energy density remains sub-Planckian but the price for it is that close to the bounce, the NEC is violated. Here the strategy is that semi-classical gravity can still be applied to analyze the theory as long as violation of NEC is not catastrophic. Since in these scenarios scale factor has to reach a minimum value and turn around, they should include an accelerated phase with $\epsilon<0$ around the bounce, followed by another accelerated phase $0<\epsilon<1$ before transiting to radiation era, with $\epsilon=2$. Note that at a regular bounce transition, $H$ goes to zero so both $R_{fr}$ and $R_{H}$ diverge. However, according to (Eq. \ref{mass}) and (Eq. \ref{RfRH})
\begin{equation}
R_f=\frac{1}{aH(2-\epsilon)},
\end{equation}
so at some later time when $H\neq 0$ and $0<\epsilon<2$, $R_{fr}$ assumes a finite value. This means after the bounce, the phase of $\epsilon<0$ which is accelerating\footnote{Note that $\frac{\ddot{a}}{a}=H^2(1-\epsilon)$ so $\epsilon<0$ is also an accelerating phase.} has a shrinking freezing radius too. In other words, the $\epsilon<0$ and $0<\epsilon<1$ phases, if long enough to produce large number of e-foldings, can lead to similar results as ordinary Inflation. 
However, the idea for these models is that the inflationary phase of the scenario is very short and fluctuations are produced in the contracting phase. Once again invoking our argument for singular bounces, modes lose e-folding numbers in the contracting phase so the amplitude can not be amplified in that phase. The only difference is that for regular bounces, modes will inevitably enter the freezing horizon close to the bounce and exit it again. During this phase, the amplitude gets amplified towards the bounce\footnote{The exact time dependence will depend on details of the model but it oscillates and gets an over all $\propto 1/a^2$ factor.} but it decays again after the bounce. So still the only way to get large amplitude would be gaining large numbers of e-foldings during this phase, which is not the desired outcome. 

Another possibility for contracting models is that one allows the existence of a dominant growing mode during the contracting phase \cite{Cai:2008qw}. There is no constraint on how low the energy density can be during the contracting phase. This means we can not exclude such a possibility based on the allowed range of the energy densities as we did for the expanding case. However, as we pointed out earlier, we expect the existence of growing infra-red tensor modes break the assumption of isotropicity for FRW metric. Note that, unlike scalar infra-red modes, the time dependent IR contribution in tensor modes have an anisotropic nature and can not be absorbed in background effect. A rigorous calculation of this effect would require computing a gauge independent and physical quantity beyond linear order (back reaction effects) while incorporating a mathematical framework such as coarse-graining to separate infra-red contributions of two point functions from the ultra-violet ones. 

Here, we just provide an intuitive argument to demonstrate that such an effect can undermine the predictions of a model. Consider a separation of the IR and UV contributions from scalar and tensor perturbations to the FRW metric in Newtonian gauge in the following way:
\begin{eqnarray}
ds^2=a^2(1+2\phi^{IR}+2\phi^{UV}) d\tau^2-a^2[(1-2\psi^{IR}-2\psi^{UV})\delta_{ij}+h^{IR}_{ij}+h^{UV}_{ij}]dx^idx^j.
\end{eqnarray}
For a local observer, the spatial variations in $IR$ modes are not distinguishable from background, so considering them as part of the background or a perturbation around it, is a matter of mathematical definition. One can absorb a constant or time-dependent $\phi^{IR}$ by rescaling time and scale factor
\begin{eqnarray}
a(\tau) &\rightarrow& \tilde{a}(\tau)\equiv a(\tau)\sqrt{1-2\psi^{IR}(\tau)} \\
d\tau &\rightarrow& d\tilde{\tau}\equiv\sqrt{1+2\phi^{IR}\over1-2\psi^{IR}}d\tau.
\end{eqnarray}
Therefore, the existence of a constant IR correction is completely benign. A growing scalar correction may deviate the FRW background from its original time dependence such that it does not remain an attractor. In this case the background metric still remains a FRW metric. One can argue that constant IR contributions from tensor modes are also benign. In this case, first we can diagonalize the spatial metric $g_{ij}$ by rotating the spatial coordinate and aligning the axes with eigenvectors of $g_{ij}$. Then we can rescale the new  coordinates to make the metric isotropic and proportional to the identity matrix:  
\begin{eqnarray}
g_{ij}=a^2[\delta_{ij}+h^{IR}_{ij}]&\xrightarrow{Rotation}& \tilde{g}_{ij}d\tilde{x}^id\tilde{x}^j=a^2[\lambda_1d\tilde{x}^1d\tilde{x}^1+\lambda_2d\tilde{x}^2d\tilde{x}^2+\lambda_3d\tilde{x}^3d\tilde{x}^3]\nonumber \\ &\xrightarrow{rescaling}& \check{g}_{ij}=a^2\delta_{ij}, ~~ d\check{x}^i=\sqrt{\lambda_i}d\tilde{x}^i.  
\end{eqnarray}
Here, we denoted eigenvalues by $\lambda_i$. This argument fails if $h^{IR}_{ij}$ is time-dependent and a growing function of time. Initially  we can pick a coordinate that satisfies the FRW criteria but soon $e_i$ vectors and $\lambda_i$ values will change. Therefore, the assumption of perturbing around FRW background is no longer a valid assumption. 



\subsection{Non standard gravitons}

At this point, a natural question is whether any possible modification of the dispersion relation for gravitons in semi-classical backgrounds could lead to enhancement of primordial tensor modes. For example, for scalar perturbations we know that a super-luminal, varying speed of sound in non-inflationary backgrounds can allow for a detectable power spectrum \cite{Bessada:2009ns,Magueijo:2002pg, Agarwal:2014ona}. Our additional knowledge of the near scale invariance of the scalar power spectrum on large scale makes it possible \cite{Geshnizjani:2011dk,Geshnizjani:2013lza} to obtain generic criteria on such models based on their dispersion relation. 
In principle, access to similar information about tensor modes could achieve the same goal. However, our argument in section \ref{sec:Bounds} was based only on the detectability threshold (amplitude for one mode) and not knowing the tilt of the spectrum. Most modifications of gravity that manifest in the dispersion relation modify the action such that the amplitude is affected too. Therefore, without knowing the tilt of the tensor power-spectrum, one has to study these models case-by-case to see how amplitude behaves. For example, Horndeski theories, \cite{Horndeski:1976gi} could lead to a varying speed of sound for tensor modes, $c_T(\tau)$. However, in these theories one could perform a disformal transformation into a frame in which speed of sound is unity \cite{Creminelli:2014wna}. Our argument from section \ref{sec:Bounds} can be easily applied in the new frame to show that the theory must either allow for super Planckian energy densities or superluminal speed of sound, in the frame with standard metric coupling to matter (i.e. ``Jordan'' frame). The steps are as follows: In the new  frame, where the new scale factor $\tilde{a}\equiv a/\sqrt{c_T}$ and the new conformal time is defined as $dy\equiv c_T(\tau) d\tau$, the action for the canonical variable $v\equiv h/\tilde{a}$ takes the following form   
\begin{equation}\label{actioncct}
S_2={1\over 2}\int d^3x~ dy \left [ v^{\prime2}- (\nabla v)^2 +\frac{\tilde{a}_{,yy}}{ \tilde{a}}v^2\right ].
\end{equation}
We can now carry out the calculations as before. The freezing transition will now occur when  
\begin{equation}
 k\simeq \frac{1}{R_{fr}}\equiv \sqrt{|\frac{\tilde{a}}{\tilde{a}_{,yy}}|}. 
\end{equation}
The size of the comoving wave numbers and predictions for power spectrum are independent of the frame. This means we can substitute the value of $k$ in the disformal frame into inequality (\ref{kahfr}) valid in the original frame (``Jordan''frame), 
\begin{equation} \label{kahfr2}
\left(\frac{k}{(a H)_{fr}}\right) \simeq \left(\frac{1}{R_{fr}(a H)_{fr}}\right) < \left(\frac{k}{(a H)_{\rm BBN}}\right)\lsim 2.5 \times 10^{-6},  
\end{equation}
where we also applied the upper bound from (\ref{epsilon150}). We can rewrite this bound as 
 \begin{equation} \label{kahfr3}
\left(\frac{1}{R_{fr}}\right) \lsim 2.5 \times 10^{-6} (a H)_{fr}.
\end{equation}
Similarly Eq. (\ref{ampfr}) can be written as 
\begin{eqnarray}\label{ampfr2}
{\cal P}^h_k  &\simeq&  \frac{1}{2}\left(\frac{k}{M_p \tilde{a}_{fr}}\right)^2\simeq  \frac{1}{2} \left(\frac{1}{M_p R_{fr} \tilde{a}_{fr}}\right)^2. 
\end{eqnarray}
Combining (\ref{kahfr3}) and (\ref{ampfr2})  we get 
\begin{eqnarray}\label{ampfr3}
{\cal P}^h_k  &\lsim&  \frac{1}{2} \left(\frac{2.5 \times 10^{-6} (a H)_{fr}}{M_p R_{fr} \tilde{a}_{fr}}\right)^2= 3.1 \times 10^{-12}\left(\frac{ c_T H^2_{fr}}{M_p^2 }\right). 
\end{eqnarray}
This shows why requiring ${\cal P}^h_k\gsim 10^{-10}$ implies either a super-Planckian energy density or superluminal propagation of gravitons 
\begin{equation}
 c_T H^2_{fr} \gsim 10~M_p^2. 
\end{equation}

A more subtle modification of dispersion relation is when speed of sound is frequency dependent $c_T=c_T(k)$. In this case, the lightcones for different wavelengths do not coincide. An example of this, a dispersion relation
of $\omega \propto k^3$, arises in Horava-Lifshitz gravity proposal \cite{Horava:2009uw} for a renormalizable theory of gravity. This theory by construction allows for superluminal propagation of waves in ultra-violet limit. An exact cosmological implementation of this theory is beyond the scope of this work. However, we can heuristically show that a detectable tensor mode requires breaking the semi-classical approximation at the level of the background as well. 
Consider $\Lambda^2_{HL}$ as the scale of Horava-Lifshitz correction such that for physical scale $k/a < \Lambda_{HL}$ we have $\omega \propto k$ and for $k/a > \Lambda_{HL}$ we have $\omega \propto k^3$. Therefore, the corrections in the action should approximately lead to a dispersion relation like
\begin{equation}
\omega_k\simeq k+\frac{k^3}{a^2\Lambda^2_{HL}}. 
\end{equation}
 By substituting $\omega_k$ in mode function (\ref{vacuum}) from above equation we can evaluate the power spectrum in the HL regime,
\begin{eqnarray}
{\cal P}^h_k&\equiv& k^3{ |v_k^2|\over M_{Pl}^2 a^2}\nonumber \\&\approx& k^3 \frac{1}{M_p^2 a^2 \omega_k}\nonumber \\&\approx&  \frac{\Lambda^2_{HL}}{M_p^2 }.
\end{eqnarray}
As we see this power spectrum is constant, scale invariant and for $\Lambda_{HL}\gsim 10^{-5} M_{p}$ can lead to ${\cal P}^h_k \gsim 10^{-10}$ \cite{Agarwal:2014ona}. If the wavelength is larger than the ordinary freezing radius of the background when the mode transition from HL to standard gravity happens (i.e. $a_{HL}(k) \sim k /\Lambda_{HL}$), the power remains frozen and can lead to possibly detectable signal. However, the bound for $\Lambda_{HL}$ implies 
\begin{equation}
 a_{HL}(k) \lsim 10^5\times \frac{k}{M_p},  
\end{equation}
which combined with (\ref{eqnkInPlanck}) leads to a lower bound on the number of  e-folding of expansion of universe at the time of setting the initial condition 
\begin{eqnarray}
 a_{HL}(k) &\lsim& 6.1 \times 10^{-55}, \\
 &\Rightarrow&
 \Delta N_{tot}\gsim 124.8. 
\end{eqnarray}
As we derived in previous section such a large number of e-foldings in non-accelerating backgrounds requires super-Planckian energy densities (\ref{eqNUB}). Therefore, while the scale of Horava-Gravity in dispersion relation does not need to be super-Planckian in order to produce large tensor perturbations,  setting adiabatic initial conditions for large scale modes requires super-Planckian energy densities.

Another interesting possibility in the realm of modified gravity theories is the existence of additional spin two degrees of freedom in some theories such as bi-metric gravity and massive gravity \cite{deRham:2014zqa}. However, generically only one of these particles mimics a massless graviton at early time. The reason is that in the end, the additional sector has to couple to the standard sector in order to produce any interesting physics. The coupling to matter makes these particles massive. 

We end this section while stressing that here we discussed a large class of scenarios inspired by quantum gravity corrections but covering all the possible proposals in the literature is beyond the scope of this article.

\section{Concluding Remarks} \label{sec:Conclusions}

In this paper we studied the theoretical consequences of detecting primordial superhorizon tensor modes in CMB. 

We started by deriving a general criterion for enhancing vacuum tensor fluctuations in expanding FRW backgrounds (section \ref{sec:WKB}). We then proved that a tensor signal with amplitude ${\cal P}_h \gsim 10^{-10}$  and the assumption of non-accelerating expansion histories ($\epsilon>1$) are not consistent within the framework of General Relativity (section \ref{sec:Bounds}). Our numerical bounds were based on possible signals of the same order of magnitude as BICEP measurements but our conclusions are valid up to few orders of magnitude smaller amplitude. We provided two different arguments for supporting our claim. First, we traced back in time the size of the wavelengths corresponding to present large scale observations, assuming $\epsilon>1$. This led to requiring freezing length being much greater than the Hubble length at freezing transition and consequently an equation of state satisfying $\epsilon \sim 2$. However, since the amplitude of tensor modes gets suppressed by a factor of $(\epsilon-2)$, an observable signal demands a Hubble parameter larger than Planck scale at the freezing transition. We also obtained a complementary bound directly from comparing the wavelength of the tensor modes today and the amount of expansion since BBN. Comparing the amount of expansion needed in order to get a sufficiently large perturbation amplitude, and the amount of expansion in a non-accelerating FRW prior to BBN, leads to energy densities of order the Planck density or larger. Considering the existence of a possible growing mode does not change this conclusion (see Appendix \ref{growing}). 

Our arguments can also be easily adjusted for many alternative scenarios which do not fall into the general relativity paradigm, such as {\it contracting} scenarios, {\it Horndeski} theories and {Horava-Lifshitz} gravity (see section \ref{sec:Alternatives}). 

We can see why in many contracting alternatives, and in the absence of a large growing mode, tensor modes cannot provide detectable signal. The argument does not apply when a significant growing mode is present but we briefly discussed why that would indicate the breakdown of FRW approximation. 

Honrndeski theories can allow for time dependent speed of tensor propagation. Applying our bounds for tensor detectability in non-accelerating backgrounds implies either a super-Planckian energy density or superluminal propagation of gravitons.

We also looked at Horava-Lifshitz gravity for the possible production of a detectable tensor signal. Our analysis suggests that while the scale of Horava-Gravity in the dispersion relation does not need to be super-Planckian in order to produce large tensor perturbations,  setting adiabatic initial conditions for large scale modes requires super-Planckian energy densities.


\section*{Acknowledgements}
The authors would like to thank Justin Khoury and Andrew Tolley for useful discussions. GG research is supported by the Discovery Grant from Natural Science and Engineering Research Council of Canada, the University of Waterloo and the Perimeter Institute for Theoretical Physics. Research at the Perimeter Institute is supported by the Government of Canada through Industry Canada and by the Province of Ontario through the Ministry of Research $\&$ Innovation. WHK is supported by the National Science Foundation under grants NSF-PHY-1066278 and NSF-PHY-1417317.
\bibliographystyle{JHEP}
\bibliography{SCINC.bib}

\appendix
\section{Bounds on magnitude of growing mode for tensor perturbations in an expanding universe} \label{growing}
We now check if the existence of a growing mode changes the conclusion obtained in Section \ref{sec:Bounds}. The size of the growing mode can be estimated by matching the solution from Eq. (\ref{vacuum}) in the limit $ k\gg 1/R_{fr}$ smoothly to the solution from Eq. (\ref{freezedsoliution}) in the limit of $k\ll 1/R_{fr}$.  More specifically, we Taylor expand $h_k$ around the freezing time for both solutions and match the leading coefficients.  
 \begin{eqnarray}
{h_k(\tau)}_{\tau \to \tau_{fr}^-}& \simeq &h_{fr}[1-(ik+(a H)_{fr})\Delta \tau],\\
{h_k(\tau)}_{\tau \to \tau_{fr}^+} &\simeq& C_1+C_2 {\Delta \tau\over a_{fr}^2}, 
\end{eqnarray}
 where $\Delta \tau=\tau-\tau_{fr}$. Comparing the coefficients we obtain
 \begin{eqnarray}
 C_1&=&h_{fr},\\
 C_2&=&-h_{fr} a_{fr}^2 (ik+(a H)_{fr}).
 \end{eqnarray}
 Substituting these coefficients for power spectrum leads to 
 \begin{eqnarray}
 {\cal P}^h_k(\tau)&\simeq &{\cal P}^h_{k}(\tau_{fr}) \left | 1-[a_{fr}^2 (ik+(a H)_{fr})\int_{\tau_{fr}}^\tau {d\tau\over a^2}]\right |^2,\\
 &\simeq& \left(\frac{k}{M_p a_{fr}}\right)^2 \left | 1-[a_{fr}^2 (ik+(a H)_{fr})\int_{\tau_{fr}}^\tau {d\tau\over a^2}]\right |^2, \label{growingpower}
 \end{eqnarray}
 where we have used (\ref{ampfr}) and note that $k$ and $(a H)_{fr}$ are related according to (\ref{kahfr}). 

 The growing mode is only significant if the expression inside the square brackets is much larger than one. To obtain an upper bound on the magnitude of that term, we note that in an expanding universe for $\tau>\tau_{fr}$, we have ${1\over a^2(\tau)}\lsim {1\over a^2_{fr}} $. This implies
 \begin{eqnarray}
 (a H)_{fr}\times \left [a_{fr}^2 \int_{\tau_{fr}}^\tau \frac{d\tau}{ a^2}\right ]\lsim   (a H)_{fr} \times \Delta \tau.
 \end{eqnarray}
  We can also obtain an upper bound on $\Delta \tau$ in the following way: Invoking the continuity equation ($\dot{\rho}/\rho=-2H\epsilon$) and Friedmann Eq. ($H^2=\rho/3M_p^2$), we get 
 \begin{eqnarray}
 \epsilon=({\sqrt{3}M_p\over 2}) \frac{-\dot{\rho}}{ \rho^{3/2}}
 \end{eqnarray}
 and then integrating it overtime and assuming $\epsilon >0$,
 \begin{eqnarray}
 && \int_{\tau_{fr}}^\tau \epsilon {d\tau\over a}= \left [{\sqrt{3}M_p \over \sqrt{\rho}}\right ]_{\tau_{fr}}^\tau\\
 &&\Rightarrow {\Delta \tau\over a(\tau)}\lsim \left ({\sqrt{3}M_p \over \epsilon_{ave}\sqrt{\rho_{fr}}}\right )\left [ {\sqrt{\rho_{fr}\over \rho(\tau)}}-1\right ],
 \end{eqnarray}
 where $\epsilon_{ave}\equiv \int \epsilon d\tau /\Delta \tau$. Combining the above inequalities and applying  Friedmann Eq. ($H^2=\rho/3M_p^2$), we can now estimate maximum contribution of the growing mode to power spectrum 

 \begin{eqnarray}
 \left|a_{fr}^2 (ik+(a H)_{fr})\int_{\tau_{fr}}^\tau {d\tau\over a^2}\right| &\lsim  & Max(1, \sqrt{2-\epsilon_{fr}}) \times a(\tau)(a H)_{fr}\left ({\sqrt{3}M_p \over \epsilon_{ave}\sqrt{\rho_{fr}}}\right )\left [ {\sqrt{\rho_{fr}\over \rho(\tau)}}-1\right ]\nonumber\\
 &\lsim  & {1\over\epsilon_{ave}(1+z)(1+z_{fr})}\left [ {\sqrt{\rho_{fr}\over \rho(\tau)}}-1\right ].\nonumber\\
 \end{eqnarray}
Note that in the last line we also used $1 \gg \sqrt{2-\epsilon_{fr}}$, since the arguments leading to  \ref{epsilon150} and \ref{epsilon30}  are still valid.  Also considering that universe is radiation dominated after BBN  and avoiding super Planckian regimes, we can impose both $z_{fr}\gsim z\gsim z_{BBN}\sim10^9$ and ${\rho_{fr}\over \rho(\tau)}\lsim {\rho_{PL}\over \rho_{BBN}}\sim 10^{88}$ to get 
 \begin{eqnarray}
  \left|a_{fr}^2 (ik+(a H)_{fr})\int_{\tau_{fr}}^\tau {d\tau\over a^2}\right|  \lsim \frac{10^{35}}{ \epsilon_{ave} (1+z_{fr})}.
 \end{eqnarray}
 However, looking back at  Eq. (\ref{growingpower}), after assuming dominance of the growing mode, substituting $k/M_P \sim 10^{-60}$ (see Eq. \ref{eqnkInPlanck}) and imposing the above upper bound,  the power can not reach close to $10^{-10}$
 \begin{eqnarray}
 {\cal P}^h_k(\tau)\lsim 10^{-50}
 \end{eqnarray}

 Note that the wavelength redshift effects and the redshift contribution from the growing mode cancel each other out. In conclusion, the above bound is too strong to allow for a growing mode leading to any observable signal.


\end{document}